\documentclass[aps,prb,notitlepage,twocolumn]{revtex4}

\usepackage{amssymb}
\usepackage{amsmath}
\usepackage{graphicx}
\usepackage{bm}
\usepackage{mdframed}
\usepackage{color}

\newcommand{\ud}{\,\mathrm{d}}

\begin{document}

\bibliographystyle{prsty}
\author{Ricardo Zarzuela, Se Kwon Kim, and Yaroslav Tserkovnyak}
\affiliation{Department of Physics and Astronomy, University of California, Los Angeles, California 90095, USA}

\title{Magnetoelectric antiferromagnets as platforms for the manipulation of solitons}

\begin{abstract}
We study the magnetic dynamics of magnetoelectric antiferromagnetic thin films, where an unconventional boundary ferromagnetism coexists with the bulk N\'{e}el phase below the N\'{e}el temperature. The spin exchange between the two order parameters yields an effective low-energy theory that is formally equivalent to that of a ferrimagnet. Dynamics of domain walls and skyrmions are analyzed within the collective variable approach, from which we conclude that they behave as massive particles moving in a viscous medium subjected to a gyrotropic force. We find that the film thickness can be used as a control parameter for the motion of these solitons. In this regard, it is shown that an external magnetic field can drive the dynamics of domain walls, whose terminal velocity is tunable with the sample thickness. Furthermore, the classification of the skyrmion dynamics is sensitive to the spatial modulation of the sample thickness, which can be easily engineered with the present (thin-film) deposition techniques. Current-driven spin transfer can trigger drifting orbits of skyrmions, which can be utilized as racetracks for these magnetic textures. 
\end{abstract}

\maketitle 

{\it Introduction.} The magnetoelectric effect refers to the induction of bulk magnetization (electric polarization) by an electric (magnetic) field.\cite{Curie-JPhys1894,Dzyaloshinskii-JETP1960,ME_exp} It requires the breaking of time-reversal symmetry, which implies the existence of a magnetic order in systems of localized spins,\cite{LLP} and of inversion symmetry (at the level of the magnetic point group).\cite{Dzyaloshinskii-JETP1960,Schmid-JPhys2008} A striking property of magnetoelectric antiferromagnets (ME-AFMs) is the emergence of an unconventional ferromagnetism at the boundaries,\cite{Binek-NatMat2010,Binek-JPhys2014} whose macroscopic signatures are well known experimentally.\cite{Binek} The existence of this boundary magnetization can be argued on symmetry grounds, since the normal $\bm{n}$ to the boundary is formally equivalent to a homogeneous electric field.\cite{Belashchenko-PRL2010}  
There is a subclass of ME-AFMs, including $\alpha$-Cr$_{2}$O$_{3}$ and Fe$_{2}$TeO$_{6}$,\cite{Binek-NatMat2010,Binek-JPhys2014,Fallarino-PRB2015} for which the magnetoelectric response is dominated by the exchange-driven mechanism and, remarkably, the boundary-induced magnetization is collinear with the (bulk) N\'{e}el order.


The ensuing ferrimagnetic state, which is described by the staggered order parameter, offers promising perspectives to manipulate the dynamics of topological solitons.\cite{TopSol} This class of magnetic textures has been intensively studied in recent years due to their topological robustness (meaning that the spin texture cannot be deformed continuously into the trivial uniform state) and to their potential use as building blocks for information storage and logic devices.\cite{TopSolInt} Of particular interest are domain walls\cite{DW} (DWs) and skyrmions\cite{Belavin-JETP1975} due to their particlelike behavior and low current threshold for skyrmion depinning.\cite{spcurrent}



\begin{figure}[t]
\begin{center}
\includegraphics[width=9cm]{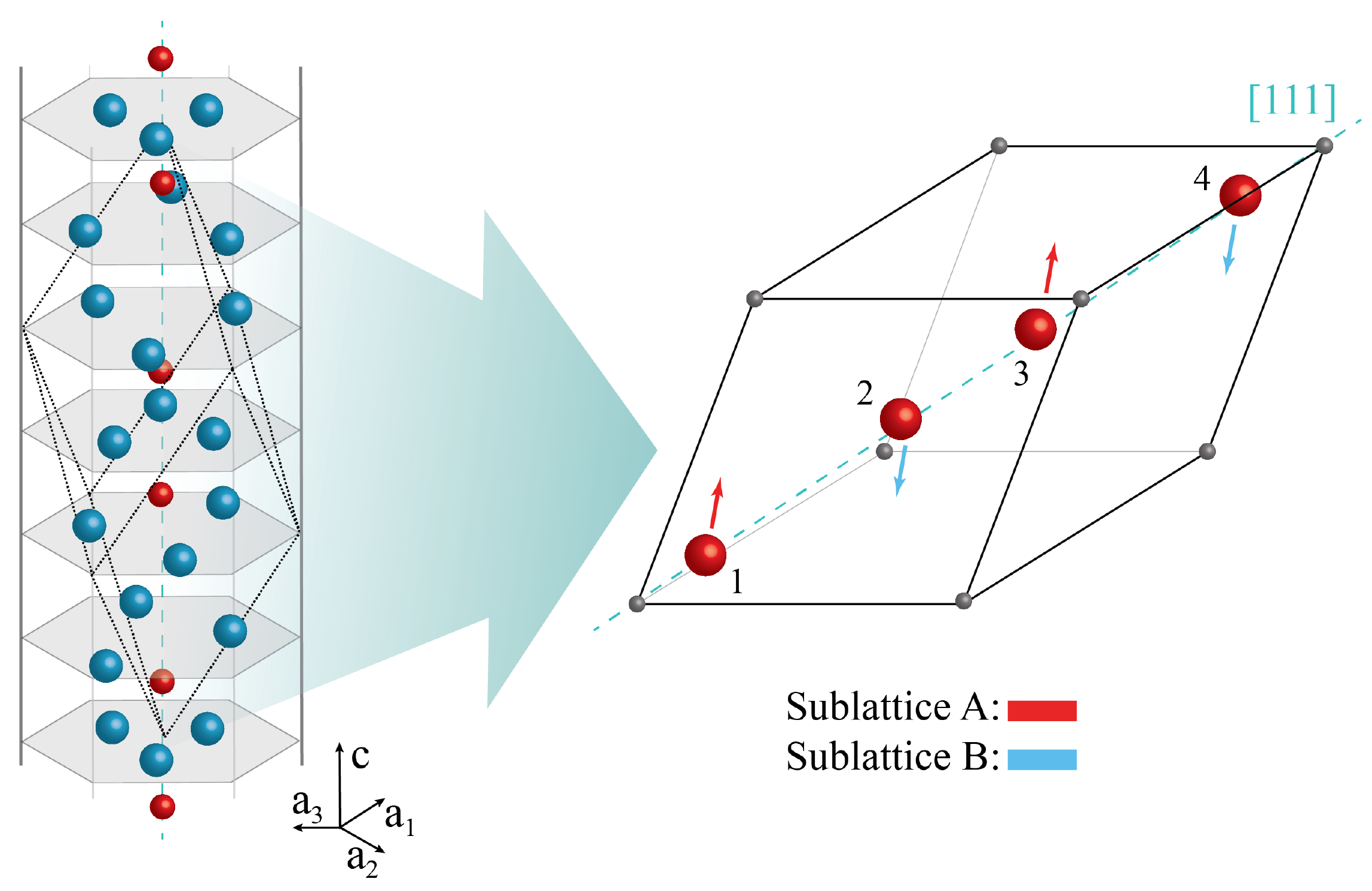}
\caption{Corundum-type crystal structure of eskolaite (mineral form of $\alpha$-Cr$_{2}$O$_{3}$). The inset depicts the corresponding unit cell. The parameters of the rhombohedral crystal lattice are $a=4.95$ \AA\,and $c=13.58$ \AA (referred to the hexagonal frame). Red [blue] spheres represent Cr$^{3+}$ [O$^{2-}$] ions. Red (sublattice A) and blue (sublattice B) arrows illustrate an spin arrangement of the Cr$^{3+}$ ions corresponding to the antiferromagnetic phase: $\bm{s}_{1}=-\bm{s}_{2}=\bm{s}_{3}=-\bm{s}_{4}$.}
\label{Fig1}
\end{center}
\vspace{-0.5cm} 
\end{figure}

In this manuscript we construct a low-energy theory for ME-AFMs with account of the aforementioned boundary effects. We focus on energy terms that favor topological solitons, with an eye on DWs and skyrmions. We furthermore study the magnetic dynamics of these two soliton classes, driven by an external magnetic field (DWs) and by an electric charge current (skyrmions), in ME-AFM thin films. 
In this regard, we consider the case of a quasi-two-dimensional (2D) ME-AFM film being subjected to spin exchange and spin-orbit coupling with a heavy metal adjacent to one of the boundaries. The motivation for this is threefold: the heavy-metal substrate (i) differentiates the two boundaries of the film, (ii) induces a Dzyaloshinskii-Moriya interaction at the interface that promotes the stabilization of skyrmion textures, and (iii) provides the medium for the charge current to flow in the insulating scenario. The DW dynamics correspond, within the collective variable approach, to that of massive particles moving in a viscous medium and subjected to a gyrotropic force depending on their precessional degree of freedom. We find that the field-driven terminal velocity of DWs shows a nonlinear behavior as a function of the sample thickness. On the other hand, the Thiele equation for skyrmions, which we derive using collective variables, is analogous to the equation of motion for a massive charged particle in a viscous medium subjected to a gyrotropic force depending on its charge. We find that these dynamics can be sustained by feasible electric currents via the spin-transfer torque effect, and that the class of skyrmion trajectories realized, including drifting orbits,\cite{Muller-PRL1992,Gerhardts-PRB} depends on the details of the film thickness profile. Our framework, albeit generic for ME-AFMs, will be built upon the example of chromia, $\alpha$-Cr$_{2}$O$_{3}$, for illustrative purposes. 

{\it Effective theory.} Chromia represents the archetypical (insulating) ME-AFM: it is a pure (bulk) antiferromagnet, meaning that it exhibits neither weak ferromagnetism\cite{Dzyaloshinskii-JETP1957,Moriya-PR1960} nor magnetic (texture) superstructures\cite{Dzyaloshinskii-JETP1964} in the ground state below the N\'{e}el temperature $T_{\textrm{N}}\simeq307$ K. It has the (bulk) symmetry of the rhombohedral space group $R\bar{3}c$ and crystallizes in a corundum-type structure, with the unit cell containing four (crystallographically equivalent) Cr$^{3+}$ ions located along a body diagonal of the rhombohedron, see Fig. \ref{Fig1}. The (low-energy) magnetic dynamics of chromia is known to correspond to that of an ordinary (bipartite) antiferromagnet,\cite{FN1} and is described by the N\'{e}el order $\bm{s}_{1}-\bm{s}_{2}+\bm{s}_{3}-\bm{s}_{4}$ and the (residual) spin density $\bm{s}_{1}+\bm{s}_{2}+\bm{s}_{3}+\bm{s}_{4}$ per unit cell.\cite{Dzyaloshinskii-JETP1957} 

\begin{figure}[t]
\begin{center}
\includegraphics[width=8.75cm]{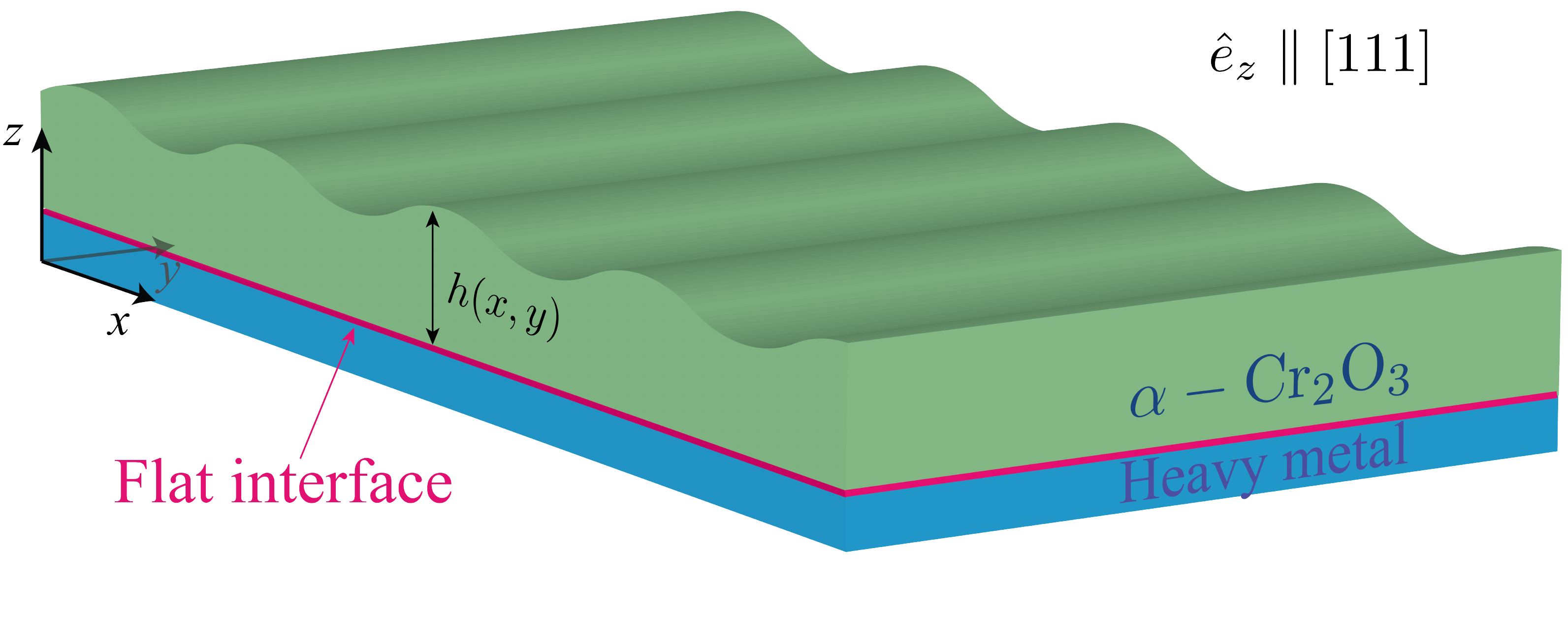}
\caption{Schematic of the heterostructure: A film made of chromia ($\alpha$-Cr$_{2}$O$_{3}$) is deposited on top of a flat heavy-metal substrate. The film is grown along the [111] direction of its rhombohedral crystal lattice and its thickness varies over the interface, which is described by the profile $h(x,y)$. This figure illustrates the example of a periodically modulated thickness along the $x$ axis.}
\label{Fig2}
\end{center}
\vspace{-0.5cm} 
\end{figure}

We consider the geometry of a chromia film deposited on top of a heavy metal, with the flat interface lying along the $(111)$ plane, see Fig. \ref{Fig2}. Our choice of coordinate system takes the $z$ axis along the trigonal axis, i.e., the normal to the interface. An equilibrium boundary magnetization emerges for this geometry since chromia exhibits a magnetoelectric response.\cite{Belashchenko-PRL2010} 
The heavy-metal substrate endows a Dzyaloshinskii-Moriya interaction in the antiferromagnetic film due to the breaking of the reflection symmetry with respect to the basal plane,\cite{Zarzuela-PRB2017} which favors spin (texture) superstructures\cite{Dzyaloshinskii-JETP1964} and, in particular, stabilizes skyrmion textures.\cite{FN2} Furthermore, it makes the two boundaries of the film become (magnetically) inequivalent and, as a result, a net boundary magnetization is present in the heterostructure. It is worth remarking that this effect on the ME-AFM is interfacial in nature, so that it will be enhanced (relative to the bulk) in thin films. 

We regard the heterostructure as a quasi-two-dimensional system along the {\it xy} plane, which we take to be isotropic at the coarse-grained level. An effective long-wavelength theory for (bulk) chromia can be developed in terms of two continuum coarse-grained fields: the (staggered) N\'{e}el order $\bm{l}$ and the normalized (volume) spin density $\bm{m}$.\cite{AFM} These fields satisfy the nonlinear local constraints $\bm{l}^{2}=1$ and $\bm{l}\cdot\bm{m}=0$, and the presence of a well-developed N\'{e}el order implies $|\bm{m}|\ll1$ on the scale of the exchange coupling. In the absence of electromagnetic fields, the corresponding three-dimensional (3D) Lagrangian density in the continuum limit becomes
\begin{equation}
\label{eq1}
\mathcal{L}_{\textrm{bulk}}[t;\bm{l},\bm{m}]=s\bm{m}\cdot(\bm{l}\times\partial_{t}\bm{l})-\frac{\bm{m}^{2}}{2\chi}-\mathcal{F}_{\textrm{stag}}[\bm{l}],
\end{equation}
to the lowest order (quadratic) in both $\partial_{t}\bm{l}$ and $\bm{m}$. Here, $s$ is the saturated (volume) spin density,\cite{FN3} $\chi$ denotes the bulk (transverse) spin susceptibility and $\mathcal{F}_{\textrm{stag}}[\bm{l}]$ stands for the total energy of the antiferromagnet.\cite{FN2} Integration out of the spin field $\bm{m}$ yields the following effective Lagrangian density for the N\'{e}el order:
\begin{equation}
\label{eq2}
\mathcal{L}_{\textrm{bulk,eff}}[t;\bm{l}]=\frac{1}{2}\chi s^{2}(\partial_{t}\bm{l})^{2}-\mathcal{F}_{\textrm{stag}}[\bm{l}],
\end{equation}
where the first term accounts for the inertia of the dynamics of the N\'{e}el order.

The boundary spin density, $\mathfrak{s}\hspace{0.03cm}\mathfrak{m}_{\textrm{b}}$, describes the spin polarized state at the interface between the chromia film and the heavy metal.\cite{FN4} Here, $\mathfrak{s}$ is the uncompensated 2D spin density\cite{FN5} and $\mathfrak{m}_{\textrm{b}}$ denotes the corresponding unit vector. It contributes to the effective theory with a 2D Lagrangian density of the form:
\begin{equation}
\label{eq3}
\mathcal{L}_{\textrm{bound}}[t,\bm{l},\mathfrak{m}_{\textrm{b}}]=\mathcal{L}_{\textrm{WZ}}[\mathfrak{m}_{\textrm{b}},\partial_{t}\mathfrak{m}_{b}]-\mathcal{F}_{\textrm{bound}}[\bm{l},\mathfrak{m}_{\textrm{b}}],
\end{equation}
where $\mathcal{L}_{\textrm{WZ}}[\mathfrak{m}_{\textrm{b}},\partial_{t}\mathfrak{m}_{b}]=-\mathfrak{s}\hspace{0.03cm}\bm{a}[\mathfrak{m}_{\textrm{b}}]\cdot\partial_{t}\mathfrak{m}_{b}$ represents the Wess-Zumino term corresponding to the 2+1D field theory of ferromagnetism, $\bm{a}[\bm{l}]$ is the vector potential for the magnetic monopole, $\nabla_{\bm{l}}\times\bm{a}=\bm{l}$,\cite{Mono} and $\mathcal{F}_{\textrm{bound}}[\bm{l},\mathfrak{m}_{\textrm{b}}]$ stands for the 2D free-energy density of the boundary magnetization. The magnetoelectric response of chromia is driven by the spin-exchange mechanism\cite{AB-AdvPhys1980,Ex_Dr_ME_Resp} at not too low temperatures; as a result, the boundary-induced magnetization $\mathfrak{m}_{\textrm{b}}$ is collinear with the staggered order parameter, $\mathfrak{m}_{\textrm{b}}\propto\bm{l}$, regardless of their orientation. 
This collinearity yields the following effective 2D Lagrangian density for the combined system\cite{Belashchenko-APL2016}
\begin{equation}
\label{eq4}
\mathcal{L}_{\textrm{eff}}[t;\bm{l}]=-\mathfrak{s}\hspace{0.03cm}\bm{a}[\bm{l}]\cdot\partial_{t}\bm{l}+\frac{\rho}{2}(\partial_{t}\bm{l})^{2}-\mathcal{F}[\bm{l}],
\end{equation}
where $\rho=\chi s^{2}h$ and $\mathcal{F}=h\mathcal{F}_{\textrm{stag}}+\mathcal{F}_{\textrm{bound}}$ are the effective inertia and (total) free-energy densities, respectively, and $h(x,y)$ denotes the 2D thickness profile of the chromia film.

The effects of an external magnetic field can be incorporated into our effective theory along the lines of Ref. \onlinecite{Andreev-SPU1980}: the net spin density, i.e. the conserved Noether charge associated with the symmetry of the Lagrangian under global spin rotations,\cite{FN7} reads $\bm{s}=\mathfrak{s}\hspace{0.03cm}\bm{l}+\rho\hspace{0.03cm}\bm{l}\times\partial_{t}\bm{l}$. In the presence of an external magnetic field $\bm{H}$, the magnetization of the chromia film can be cast as $\bm{M}=g\,\mathfrak{s}\hspace{0.03cm}\bm{l}+g\rho\hspace{0.03cm}\bm{l}\times\partial_{t}\bm{l}+\hat{\chi}^{\star}\bm{H}$, where $g$ denotes the gyromagnetic ratio and $\hat{\chi}^{\star}$ is the magnetic susceptibility tensor. Since $\bm{M}=\partial\mathcal{L}_{\textrm{eff}}/\partial\bm{H}$, the susceptibility must take the form $\chi_{ij}^{\star}=\rho g^{2}(1-l_{i}l_{j})$ and, therefore, the effective Lagrangian density is extended to
\begin{equation}
\label{eq5}
\mathcal{L}_{\textrm{eff}}[t;\bm{l}]=-\mathfrak{s}\hspace{0.03cm}\bm{a}[\bm{l}]\cdot\partial_{t}\bm{l}+\frac{\rho}{2}(\partial_{t}\bm{l}-g\bm{l}\times\bm{H})^{2}-\mathcal{F}[\bm{l}],
\end{equation}
where $\mathcal{F}[\bm{l}]$ includes the Zeeman term, $-g\mathfrak{s}\hspace{0.03cm}\bm{l}\cdot\bm{H}$. To conclude, a phenomenological approach well suited to incorporate dissipation into the combined system considers a Gilbert-Rayleigh dissipation function, $\mathcal{R}[\bm{l}]=hs\alpha(\partial_{t}\bm{l})^{2}/2$, which is half of the dissipation power density. Here, $\alpha$ denotes the bulk Gilbert-damping constant and we have omitted the boundary contribution to dissipation.\cite{FN8} Henceforth we will treat chromia as a ferrimagnet and study the dynamics of DWs and skyrmions, with an eye on the thickness acting as a control parameter. Note that this approach differs  from previous studies based on the thermal and/or chemical control of the saturated spin density.\cite{Kim-PRB2017}

{\it Domain wall dynamics.}  We restrict ourselves, in what follows, to magnetic spin textures whose dynamics are encoded in the time evolution of a discrete set of soft modes. Of particular interest are DWs,\cite{DW} which in the low-frequency regime can be described by their center of mass $X$ and their azimuthal angle $\Phi$ within the collective variable approach. Let $x$ denote the direction of propagation of the DW and the thickness $h$ be uniform. With account of the ansatz $\cos\Theta(x)=\tanh[(x-X)/\delta]$ for the out-of-plane component of the N\'{e}el order in the spherical-coordinate representation, $\bm{l}=(\sin\Theta\cos\Phi,\sin\Theta\sin\Phi,\cos\Theta)$, the Euler-Lagrange equations for the Lagrangian density \eqref{eq4} become\cite{Belashchenko-APL2016}
\begin{align}
\label{eq6}
2\delta\mathfrak{s}\dot{\Phi}+2\rho\ddot{X}+2s\alpha h\dot{X}=&\delta F_{X},\\
\label{eq7}
-2\mathfrak{s}\dot{X}+2\delta\rho\ddot{\Phi}+2\delta s\alpha h\dot{\Phi}=&F_{\Phi},
\end{align}
where $F_{X}=-\delta_{X}F$ and $F_{\Phi}=-\delta_{\Phi}F$ are the thermodynamic forces conjugate to $X$ and $\Phi$, respectively, with $F$ being the total free energy. Here, the DW width is given by $\delta=\sqrt{A/|K|}$, with $A$, $K$ being the exchange stiffness and anisotropy constants, respectively.\cite{FN2} 

\begin{figure}[ht]
\begin{center}
\includegraphics[width=0.95 \columnwidth]{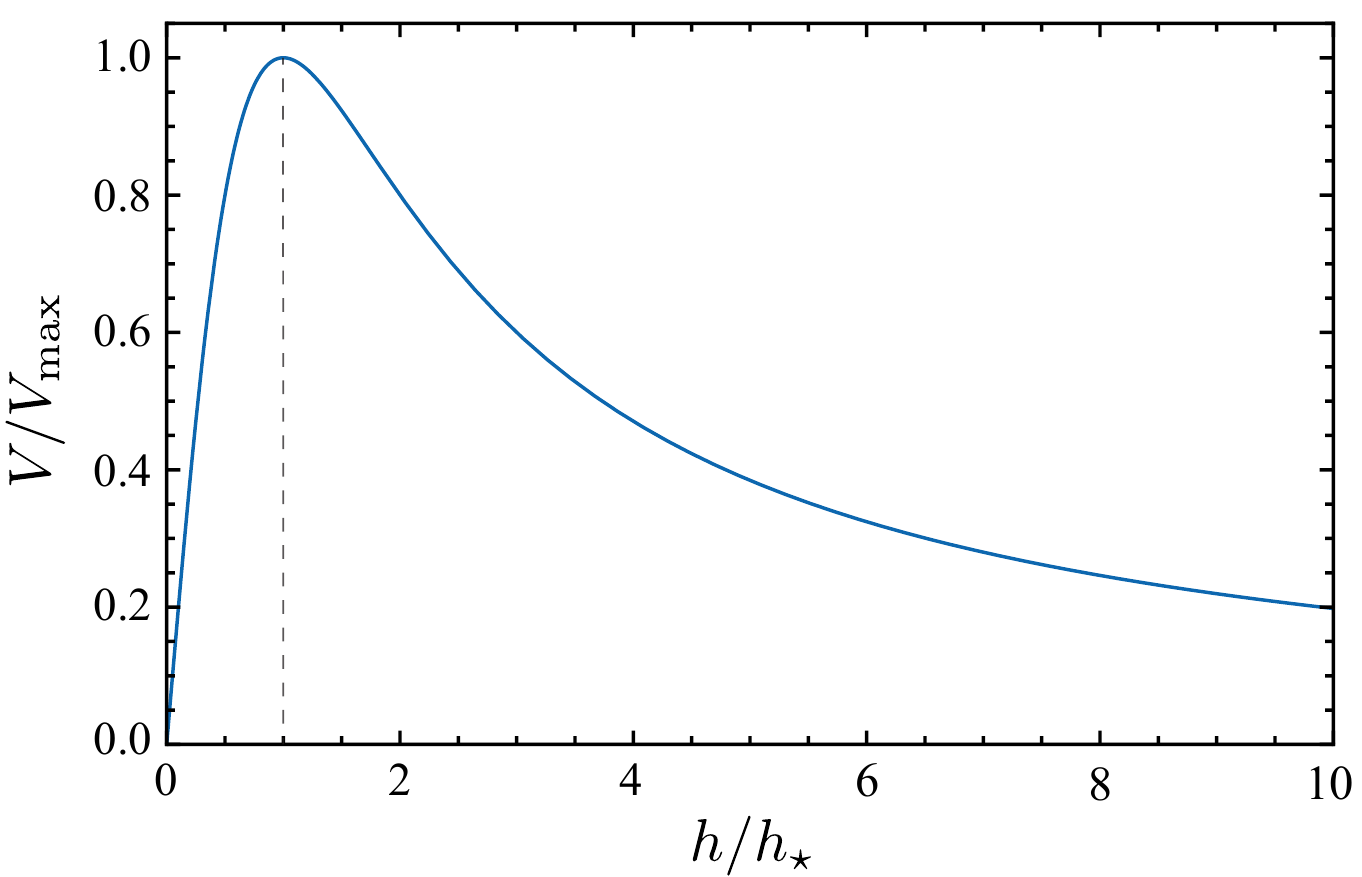}
\caption{Dependence of the field-driven terminal velocity of the domain wall on the thickness of the ME-AFM film. Both quantities have been normalized to the maximum velocity $V_{\textrm{max}}=g\delta H_{z}/2$ and the length scale $h_{\star}=\mathfrak{s}/s\alpha$, respectively. The dashed line illustrates the maximum of the terminal velocity reached at the value $h=h_{\star}$ of the film thickness.}
\label{Fig3}
\end{center}
\vspace{-0.2cm} 
\end{figure}

In the presence of a strong magnetic field, $\bm{H}=H_{z}\hat{e}_{z}$, the energetics of the ME-AFM are dominated by the Zeeman coupling, so that the thermodynamic forces can be approximated by $F_{X}\simeq -2g\mathfrak{s}H_{z}$ and $F_{\Phi}\simeq0$. Eq. \eqref{eq7} therefore dictates that $\dot{\Phi}|_{\textrm{st}}=\mathfrak{s}(\dot{X}|_{\textrm{st}})/\delta s\alpha h$ is the angular velocity of the DW in the steady state. 
By substituting it into Eq. \eqref{eq6} we obtain the following expression for the field-driven terminal velocity of the DW,
\begin{equation}
\label{eq8}
V=\frac{2h/h_{\star}}{1+(h/h_{\star})^{2}}V_{\textrm{max}},
\end{equation}
where $h_{\star}=\mathfrak{s}/s\alpha$ and $V_{\textrm{max}}=g\delta H_{z}/2$ is the maximum velocity. Its reduction as compared to $V_{\textrm{max}}$ is due to the ferromagnetic nature of the boundaries and $h/h_{\star}$ parametrizes the effective damping $\alpha_{\textrm{eff}}$. Since the usual DW terminal velocity goes as $\propto\alpha_{\textrm{eff}}/(1+\alpha_{\textrm{eff}}^{2})$,\cite{DW} we obtain a maximum at the value $h=h_{\star}$ of the sample thickness, see Fig. \ref{Fig3}.


{\it Skyrmion dynamics.} 
Skyrmions represent the epitome of spatially localized solitons in two dimensions,\cite{Belavin-JETP1975} are topologically charged and arise in magnetic systems with spin-orbit coupling.\cite{Skyrmion} Within the collective variable approach, these spin textures can be described by their center of mass, $\bm{X}=(X,Y)$, in the low-frequency regime,\cite{FN9} i.e. $\bm{l}[t,\bm{r}]=\bm{l}_{0}[\bm{r}-\bm{X}(t)]$. With account of this ansatz for the order parameter, the Euler-Lagrange equations for the Lagrangian density \eqref{eq4} now become
\begin{equation}
\label{eq9}
\rho_{M}h(\bm{X})\ddot{\bm{X}}+4\pi\mathfrak{s}\mathcal{Q}\hspace{0.03cm}\dot{\bm{X}}\times\hat{e}_{z}+\Gamma h(\bm{X})\dot{\bm{X}}=\bm{F}_{\textrm{int}}+\bm{F}_{J},
\end{equation}
where the terms on the left-hand side represent (from left to right) the inertial, Magnus and friction forces acting on the skyrmion, respectively. Here, $\rho_{M}=\chi s^{2}\int_{\mathbf{R}^{2}}\ud x\ud y (\partial_{x}\bm{l}_{0})^{2}$ is the mass density (per thickness) of the skyrmion texture, $\Gamma=\alpha\rho_{M}/\chi s$ denotes the viscous coefficient, and $\mathcal{Q}=\int_{\mathbf{R}^{2}}\ud x\ud y\hspace{0.1cm}\bm{l}_{0}\cdot(\partial_{x}\bm{l}_{0}\times\partial_{y}\bm{l}_{0})/4\pi$ is the Pontryagin index (so-called topological charge) of the skyrmion texture, which is a topological invariant and provides a measure of the wrapping of the order parameter $\bm{l}_{0}(\bm{r})$ around the unit sphere. 

Our Thiele equation\cite{Thiele-PRL1973} for the soft modes of the texture, Eq. \eqref{eq9}, is derived for the case of thickness profiles $h(x,y)$ smooth over length scales larger than the typical size of the skyrmion. This requirement translates into the adiabatic condition $|\partial_{x,y}\ln h|R_{\star}\ll 1$, where $R_{\star}$ denotes the skyrmion radius.\cite{FN2} Finally, $\bm{F}_{\textrm{int}}=-\delta_{\bm{X}}F$ is the internal force and $\bm{F}_{J}$ represents the force exerted on the skyrmion by a charge current $\bm{J}$ flowing in the heavy-metal substrate. The latter originates in the spin-transfer torque\cite{Hals-PRL2011} exerted on the spin texture by the applied charge current via the (exchange) proximity effect, and takes the form $F_{J,i}:=\int_{\bm{R}^{2}}\ud x\ud y\left\{\zeta_{1}\hspace{0.03cm}\bm{l}_{0}\cdot\left[(\bm{J}\cdot\nabla)\bm{l}_{0}\times\partial_{i}\bm{l}_{0}\right]-\zeta_{2}\hspace{0.03cm}\partial_{i}\bm{l}_{0}\cdot(\bm{J}\cdot\nabla)\bm{l}_{0}\right\}$, $i=x,y,z$, where $\zeta_{1}$ and $\zeta_{2}$ are the phenomenological constants for the reactive and dissipative components of the (texture-induced) spin-transfer torque,\cite{Kim-PRB2017} respectively. Note that spin-orbit torques do not exert an effective force on the skyrmion texture for the rigid (skyrmion) ansatz.\cite{Lin-2017} Henceforth we will assume the lowest energy configuration for the skyrmion textures (corresponding to $\mathcal{Q}=\pm1$),\cite{FN2} and focus on their current-driven dynamics. 


We consider the simple scenario of one-dimensional (1D) thickness profiles (along a direction defined as the $x$ axis), $h\equiv h(x)$. A first insight into the ensuing skyrmion dynamics can be gained by analyzing the case of absence of dissipation and external forces, $\Gamma=0,\,\bm{F}_{\textrm{int}}+\bm{F}_{J}=\bm{0}$: Since the equation of motion \eqref{eq9} is formally equivalent to that for massive 2D charged particles subjected to a (spatially modulated) transversal magnetic field,\cite{FN10} we obtain the constants of motion
\begin{align}
\label{eq10}
\frac{\Pi_{y}}{\rho_{M}}&=\dot{Y}-\int_{0}^{X}\ud z\,\omega_{0}(z),\\
\label{eq11}
\frac{\mathcal{F}_{\textrm{sky}}}{\rho_{M}}&=\frac{1}{2}\left[\dot{X}^{2}+\left(\frac{\Pi_{y}}{\rho_{M}}+\int_{0}^{X}\ud z\,\omega_{0}(z)\right)^{2}\right],
\end{align}
representing the $y$-component of the canonical-momen\-tum and free-energy densities (normalized to the mass density), respectively. Here, $\omega_{0}(X)=4\pi\mathfrak{s}\mathcal{Q}/\rho_{M}h(X)$ is the (spatially modulated) analog of the cyclotron frequency. Skyrmion trajectories can be obtained by integrating this system of first-order invariants; drifting orbits are expected to occur for monotonically decreasing and periodically modulated profiles.\cite{Muller-PRL1992,Gerhardts-PRB} Dissipation results in a loss of energy for skyrmions, which translates into a decrease of the amplitude of the periodic $x$-motion and of the drift velocity $\langle \dot{Y}\rangle$ with time. The current-induced force, however, can be chosen to compensate this effect and to stabilize these skyrmion trajectories.

\begin{figure}[ht]
\begin{center}
\includegraphics[width=8.7cm]{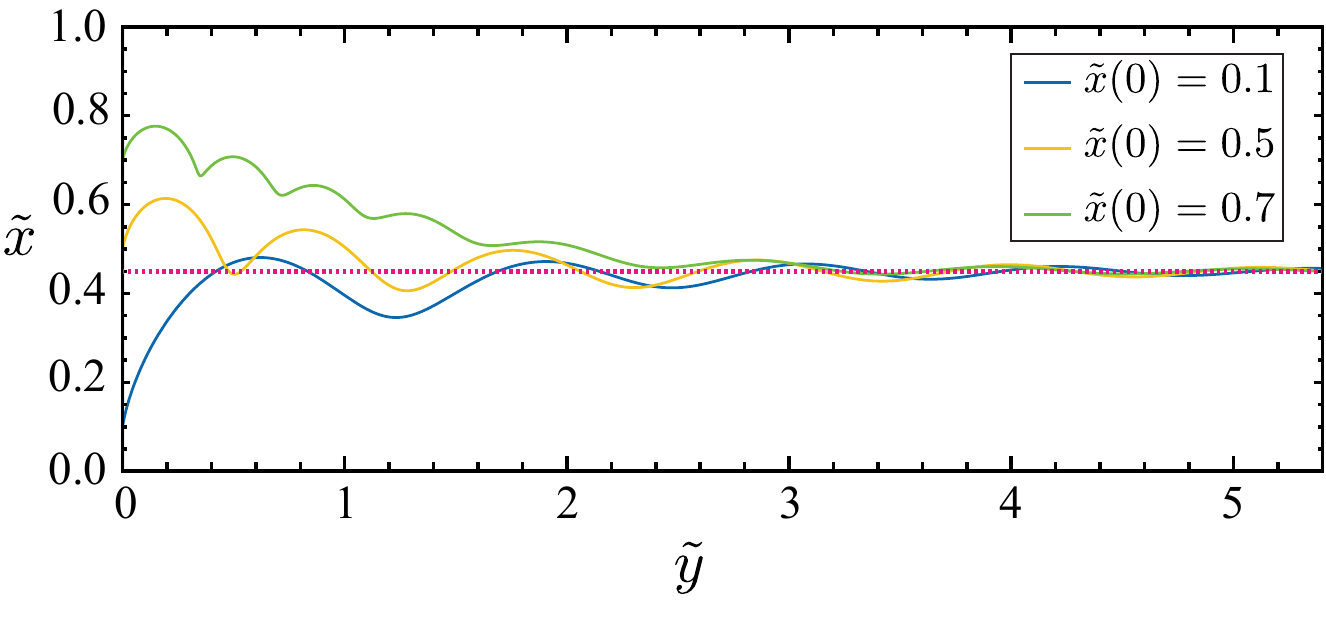}
\caption{Drifting-like orbits of skyrmions with topological charge $\mathcal{Q}=1$ subjected to the current-induced force $\tilde{\bm{F}}_{J}=0.8(4\pi\hat{e}_{x}+\mathcal{K}\hat{e}_{y})$ for the hyperbolic thickness profile $h(\tilde{x})/\mathcal{T}=1+0.7\tanh[10(\tilde{x}_{c}-\tilde{x})]$, with $\tilde{x}_{c}=0.45$. These trajectories are calculated by numerical integration of the dimensionless equations of motion \eqref{eq13} with the following values of the parameters: $\mathcal{K}=0.9\pi$ and $4\pi\mathfrak{s}/\alpha s\mathcal{T}=4\pi$. In the calculations we have taken the initial velocity $\dot{\tilde{x}}(0)=\dot{\tilde{y}}(0)=0$, and initial position along the $x$ axis ($\tilde{y}(0)=0$): $\tilde{x}(0)=0.1$ (blue), $\tilde{x}(0)=0.5$ (yellow) and $\tilde{x}(0)=0.7$ (green). The magenta dashed line at $\tilde{x}=\tilde{x}_{c}$ depicts the (attractive) racetrack for the skyrmion dynamics.}
\label{Fig4}
\end{center}
\vspace{-0.5cm} 
\end{figure}

As a specific illustrative example, we study the linear profile $h(x)=\mathcal{T}-h_{0}x/L$, where $L$ is the lateral size of the chromia film (spanning the domain $0\leq x\leq L$) and the heights $h_{0},\mathcal{T}$ satisfy the conditions $h_{0}<\mathcal{T}$ and $h_{0}R_{\star}/L\ll\mathcal{T}$. In the steady state, solutions of Eq. \eqref{eq9} are given by
\begin{equation}
\label{eq12}
\dot{\bm{X}}=\frac{1}{16\pi^{2}\mathfrak{s}^{2}+\Gamma^{2}h^{2}(X)}\begin{pmatrix}
\Gamma h(X) F_{J,x}-4\pi\mathfrak{s}\mathcal{Q} F_{J,y}\\
4\pi\mathfrak{s}\mathcal{Q} F_{J,x}+\Gamma h(X) F_{J,y}
\end{pmatrix}.
\end{equation}

Let us now apply a current-induced force $\bm{F}_{J}\propto4\pi\mathfrak{s}\mathcal{Q}\hat{e}_{x}+$ $\Gamma h(X_{c})\hat{e}_{y}$,\cite{FN11} parametrized by a certain intermediate position $0<X_{c}<L$. 
The components of the terminal velocity \eqref{eq12} read $V_{x}\propto4\pi\mathfrak{s}\mathcal{Q}\,\Gamma[h(X)-h(X_{c})]$ and $V_{y}>0$.\cite{FN12} Therefore, since $V_{x}(X\hspace{0.1cm}{\scriptstyle\lessgtr}\hspace{0.1cm} X_{c})\hspace{0.05cm}{\scriptstyle\gtrless}\hspace{0.07cm}0$, the line $x=X_{c}$ becomes an attractor for the dynamics of skyrmions with a topological charge $\mathcal{Q}$. The linear case illustrates the following general statement: {\it given any 1D thickness profile monotonically decreasing along the} ({\it so-defined}) {\it $x$ axis, we can generate a self-focusing skyrmion racetrack transversal to any $x$-coordinate by tuning the current-induced force} ({\it or, equivalently,\cite{FN11} the current density in the substrate}).

We illustrate this statement by performing the numerical calculation of skyrmion trajectories in the $XY$ plane. Fig. \ref{Fig4} depicts the generation, for a hyperbolically decreasing thickness profile, of a self-focusing skyrmion racetrack sustained by the appropriate current-induced force. The numerical trajectories are obtained by integrating the dimensionless form of Eq. \eqref{eq9}:
\begin{equation}
\label{eq13}
\mathcal{K}\,\frac{h(\tilde{X})}{\mathcal{T}}\left[\frac{d^{2}\tilde{\bm{X}}}{d\tilde{t}^{2}}+\frac{d\tilde{\bm{X}}}{d\tilde{t}}\right]+\frac{4\pi\mathfrak{s}\mathcal{Q}}{\alpha s\mathcal{T}}\frac{d\tilde{\bm{X}}}{d\tilde{t}}\times\hat{e}_{z}=\tilde{\bm{F}}_{J},
\end{equation}
where the space and time are rescaled with respect to the lateral size $L$ and the relaxation time $\tau=\chi s/\alpha$, respectively, and $\mathcal{K}=\int_{\mathbf{R}^{2}}\ud x\ud y\hspace{0.1cm}(\partial_{x}\bm{l}_{0})^{2}$ denotes a (dimensionless) geometric factor determined by the skyrmion texture.

{\it Discussion.} We have shown that ME-AFMs offer an attractive platform to control (fast) antiferromagnetic dynamics of DWs driven by an external magnetic field, as in the ferrimagnetic counterparts.\cite{Ferr-DW} Similar dynamics could be also triggered by an applied charge current, which exerts a force on the DW via the spin-transfer effect. The latter contributes to the equations of motion \eqref{eq6}-\eqref{eq7} with two components $F_{J,X}$ and $F_{J,\Phi}$ to the total force, respectively.\cite{Tatara-PRL2004} Therefore, the expression \eqref{eq8} for the terminal velocity is still valid upon redefinition of the maximum velocity, $V_{\textrm{max}}(h)=[\delta F_{J,X}-(h_{\star}/h)F_{J,\Phi}]/4\mathfrak{s}$, which now becomes thickness dependent. It is important to mention that the ferromagnetism emerging in Ref. \onlinecite{Belashchenko-APL2016} is a bulk property of chromia, which is controlled by an external electric field. As a result, the corresponding field-driven terminal velocity of the DW is insensitive to the sample thickness, unlike the present case.


Regarding skyrmions, the theory presented in this manuscript is, in a way, complementary to that of Ref. \onlinecite{Kim-PRB2017} for ferrimagnets, since both share the same Thiele equation for the dynamics of skyrmions but have different control variables: In our case, the thickness profile plays this role through the inertia and the viscous coefficient, whereas in the ferrimagnetic case it is given by the saturated spin density of the system.  That being said, our framework for the manipulation of skyrmion textures can be more advantageous for several reasons: First, from an engineering perspective, an accurate shaping of the sample surface is more feasible than the thermal or chemical control of the saturated spin density required in Ref. \onlinecite{Kim-PRB2017}. Second, ferrimagnetic materials behave effectively as ferromagnets in almost all circumstances, the only exception being when (a region of) the system is driven into the (angular-momentum) compensation point, where they exhibit an antiferromagnetic behavior. On the contrary, bulk ME-AFMs are intrinsically antiferromagnetic, with the ferrimagnetic character emerging in the so-called holographic fashion\cite{Holo} (it is encoded in the boundaries of the system); the ensuing dynamics are, therefore, suitable to be exploited in the context of antiferromagnetic spintronics. Furthermore, the exchange-driven collinearity between the boundary magnetization and the bulk N\'{e}el order allows the imaging of (the dynamics of) antiferromagnetic textures by means of magneto-optical techniques.

\acknowledgements
We thank O. Tchernyshyov for insightful remarks and P. Upadhyaya for drawing our attention to this class of magnetic materials. This work has been supported by NSF-funded MRSEC under Grant No. DMR-1420451 and by the Army Research Office under Contract No. W911NF-14-1-0016. R.Z. thanks Fundaci\'{o}n Ram\'{o}n Areces for support through a postdoctoral fellowship within the XXVII Convocatoria de Becas para Ampliaci\'{o}n de Estudios en el Extranjero en Ciencias de la Vida y de la Materia.


\begin{thebibliography}{99}
\bibitem{FN3} It reads $s=\hbar S/\mathcal{V}$, where $S$, $\mathcal{V}$ are the (dimensionless) spin and volume per site, respectively.
\bibitem{FN4} This coarse-gained field may be defined (modulo normalization) by $\sum_{\bm{r}_{i}\in\mathcal{S}}\bm{S}_{i}\delta^{(2)}(\bm{r}-\bm{r}_{i})$, where $i$ runs over all (lattice) sites belonging to the interface $\mathcal{S}$ between the film and the heavy metal, and $\bm{r}_{i}$, $\bm{S}_{i}$ are the position and spin vectors at the $i$-th site, respectively.
\bibitem{FN8} We focus on this case for the sake of simplicity. In general, we should include a boundary contribution $\mathfrak{s}\alpha_{\textrm{b}}(\partial_{t}\bm{l})^{2}/2$ to the Gilbert-Rayleigh dissipation function, with $\alpha_{\textrm{b}}$ being the boundary Gilbert-damping constant. We can account for this term by redefining the thickness profile function as $h\Rightarrow h'=h+\mathfrak{s}\alpha_{\textrm{b}}/s\alpha$ in Eqs. \eqref{eq8} and \eqref{eq12}.
\bibitem{FN9} In the low-frequency regime, the coordinates encoding details of the geometry (shape) correspond to hard modes of the texture. Therefore, we can take the skyrmions to be rigid in the first approximation.
\bibitem{FN10} The following correspondences must be accounted for: $m_{e}\rightarrow\rho_{M}$ and $\bm{B}(X)=[4\pi\mathfrak{s}/h(X)]\,\hat{e}_{z}$.
\bibitem{FN11} For a constant (and therefore divergenceless) charge current, $\bm{J}=J_{x}\hat{e}_{x}+J_{y}\hat{e}_{y}$, we can cast the current-induced force $\bm{F}_{J}$ as 
\begin{equation}
\begin{pmatrix}
F_{J,x}\\
F_{J,y}
\end{pmatrix}=
\begin{pmatrix}
-\frac{\rho_{M}}{\chi s^{2}}\zeta_{2}&-4\pi\mathcal{Q}\zeta_{1}\\
4\pi\mathcal{Q}\zeta_{1} & -\frac{\rho_{M}}{\chi s^{2}}\zeta_{2}
\end{pmatrix}
\begin{pmatrix}
J_{x}\\
J_{y}
\end{pmatrix}.\nonumber
\end{equation}
Since the determinant of this linear system is nonzero, there is a unique charge current $\bm{J}$ inducing the force $\bm{F}_{J}\propto4\pi\mathfrak{s}\mathcal{Q}\hat{e}_{x}+$ $\Gamma h(X_{c})\hat{e}_{y}$ proposed in the main text.
\bibitem{FN12} We are implicitly assuming that the constant of proportionality appearing in the expression for $\bm{F}_{J}$ is positive. As a result, the sign of $V_{x}$ corresponds to that of $h(x)-h(x_{c})$.
\end{thebibliography}
\end{document}